\documentclass[12pt,twoside]{article}
\usepackage{fleqn,espcrc1}

\usepackage{graphicx}
\usepackage[figuresright]{rotating}

\newcommand{\AmS}{{\protect\the\textfont2
  A\kern-.1667em\lower.5ex\hbox{M}\kern-.125emS}}

\title{The QGP-Stall or Not ?}

\author{B.R. Schlei\address{Theoretical Division, T-1, 
        Los Alamos National Laboratory, \\ 
        P.O. Box 1663, MS-E541, Los Alamos, NM 87545, U.S.A.}%
        \thanks{This work has been supported by the U.S. Department of 
        Energy.}}
       
\begin{document}

\maketitle

\begin{abstract}
        \vspace*{-4.0cm}
        \hspace*{\fill}{\normalsize LA-UR-99-3620} \vspace*{3.5cm}\\
In 1996, D.H. Rischke and M. Gyulassy \cite {rischke96} proposed to use the 
time-delay signal in Bose-Einstein correlations (BEC) as a signature for 
the possible formation of a quark-gluon plasma (QGP) in relativistic 
heavy-ion collisions.
In particular, they suggested to measure the ratio of the transversal
interferometry radii, $R_{out}/R_{side}$, which can be obtained by
fitting experimental BEC data with a parametrization introduced by
G. Bertsch \cite{bertsch88} et al. in 1988. The transverse radius 
parameter $R_{out}$ has compared to the transverse radius parameter 
$R_{side}$ an additional temporal dependence which should be sensitive to 
a prolonged lifetime of a fireball, in case a QGP was formed in a 
relativistic heavy-ion collision. This latter phenomenon is known under the 
term ``QGP stall''.
In the following, I shall explain why I believe that the time-delay signal 
in Bose-Einstein correlations {\it is not} a good signature for the possible 
formation of a QGP.
\end{abstract}

\section{The Framework And The QGP Stall at CERN/SPS Beam Energies}

Let us consider two quite different equations of state (EOS) of nuclear
matter within a {\it true} relativistic hydrodynamic framework 
(i.e., HYLANDER-C) \cite{schlei97}. The first one, EOS-I, has a 
phase-transition to a QGP at $T_C$ = 200 $MeV$ ($\epsilon_C$ = 1.35 
$GeV/fm^3$) \cite{schlei99a}. The second EOS, EOS-II, is a purely hadronic EOS, 
which has been extracted from the transport code RQMD (cf., Ref. 
\cite{schlei98}) under the assumption of fully acchieved thermalization. 
If one assumes for each EOS {\it different} initial conditions before the
hydrodynamical expansions, one can fit simultaneously hadronic single 
inclusive momentum spectra and BEC, which have been measured recently by 
the CERN/NA44 and CERN/NA49 Collaborations (cf., \cite{schlei99a,schlei98}), 
respectively. 
In particular, for the acceptance of the NA44 experiment a ratio 
$R_{out}/R_{side} \approx$ 1.15 was found \cite{schlei99a} while 
using both EOS, EOS-I and EOS-II. Little difference was seen in the BEC of 
identical pion pairs while considering the two different EOS.

\section{The QGP Stall At BNL/RHIC Beam Energies And Conclusions}

In the following, we shall assume for central Au+Au collisions at BNL/RHIC 
beam energies a set of fireball initial conditions, IC-I, which are similar
to those as described in Ref. \cite{schlei99c}. From these fireball initial 
conditions, IC-I, single inclusive momentum spectra have been calculated 
while using EOS-I and EOS-II in the hydrodynamic expansions. We note, that 
the rapidity spectra of both calculations differ in width and nomalization 
significantly \cite{schlei99b}. 
Fig. 1. shows, that the isothermes of the transversely expanding fluids at 
longitudinal coordinate $z=0$ also differ significantly. Since there will be 
only one set of measured data, we shall fit the calculation using EOS-II to 
the single inclusive momentum spectra of the calculation using EOS-I. 
In doing so, we find new initial conditions, IC-II \cite{schlei99b}. 
But now the space-time picture of the evolving fireball at freeze-out is 
again very similar to the one using EOS-I with IC-I.
\begin{figure}[t]
\includegraphics*[scale=1.0]{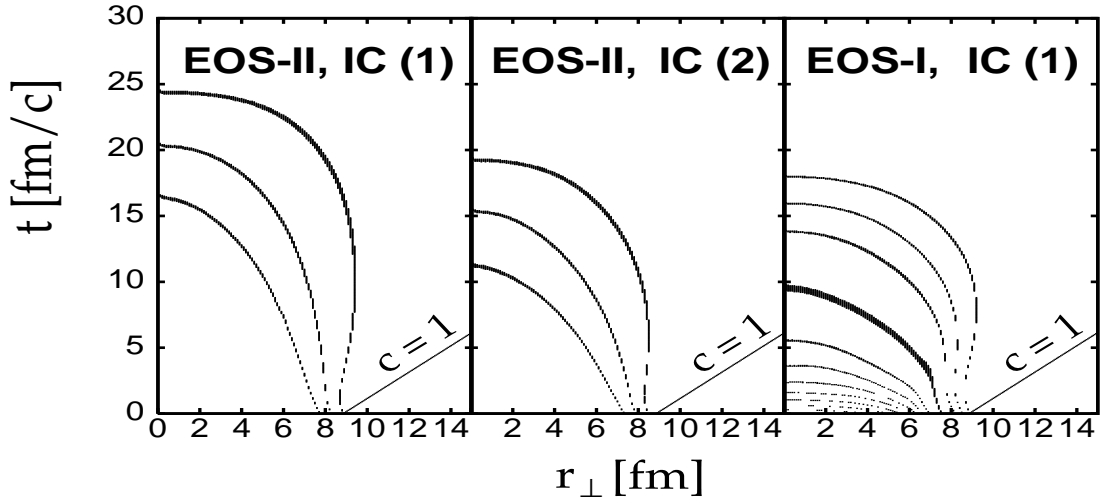}
\caption{Isothermes in steps of $\Delta T = 20MeV$. The outer contours equal
$T_f = 140MeV$.}
\end{figure}

If one calculates BEC of identical pion pairs or identical kaon pairs,
resepectively, one finds \cite{schlei99b} while comparing the calculations 
using EOS-I with IC-I and EOS-II with IC-II, respectively, {\it no} 
significant differences in the extracted ratios $R_{out}/R_{side}$ regardless 
of the pair kinematics under consideration. In particular, the assumption of 
the PHENIX detector acceptance \cite{schlei99b} leads to a ratio 
$R_{out}/R_{side} \approx$ 1.65 for both choices of EOS.

In summary, the larger ratio $R_{out}/R_{side}$ at RHIC beam energies
appears to be rather a consequence of the expected higher energy deposit
in the fireball during the heavy-ion collision, but it appears {\it not} to
be an indicator of the present or absent phase-transition to a QGP.
Of course, more theoretical analysis is neccessary, but there is strong 
evidence, that BEC do not provide a good QGP signature, since we do not
understand the initial state of a heavy-ion collision well enough yet.

\end{document}